\documentclass[aps,prb,twocolumn,superscriptaddress]{revtex4}
\usepackage{ae}
\usepackage[T1]{fontenc}
\usepackage[ansinew]{inputenc}
\usepackage{amsmath}
\usepackage{amssymb}
\usepackage{graphicx}
\usepackage{color}
\usepackage[colorlinks]{hyperref}
\usepackage{epstopdf}

\begin{document}

\title{Topological instability of two-dimensional conductors}

\author{A.M. Kadigrobov$^{*}$}
\affiliation{Theoretische Physik III, Ruhr-Universitaet Bochum, D-44801 Bochum, Germany}
\affiliation{Department of Physics, Faculty of Science, University of Zagreb, Croatia}
\author{A. Bjeli\v{s}}
\affiliation{Department of Physics, Faculty of Science, University of Zagreb, Croatia}
\author{D. Radi\'{c}$^{\dag}$}
\affiliation{Department of Physics, Faculty of Science, University of Zagreb, Croatia}

\begin{abstract}
In the present paper we propose a mechanism of the structural instability with a periodic charge ordering
in two-dimensional isotropic conductors with a closed Fermi surface which completely excludes the conventional
nesting mechanism. We show that the structural instability in such conductors may arise as a topological
reconstruction under which the initially closed Fermi surface is transformed into an open one. We have found
that the order parameter of the charge ordering ground state  may exceed one hundredth of the Fermi energy.
Furthermore, this charge ordering is a quantum phase transition with respect to the dimensionless coupling
constant $\lambda$ related to the mechanism that drives the band reconstruction (e. g. electron-phonon coupling), with the critical value given by $\lambda_c= (1+2/\pi)^{-1}$. Preliminary estimations show
that the suggested mechanism  can be the origin of density waves observed in such materials as high$-T_c$
cuprates or graphite intercalates.
\end{abstract}

\maketitle

\let\thefootnote\relax\footnotetext{e-mail: $^{*}kadig@tp3.rub.de$, $^{\dag}dradic@phy.hr$}

\textbf{I. Introduction}

\bigskip

Almost ninety years ago Peierls \cite{Peierls}  predicted an instability of one-dimensional metals with respect to
the spontaneous arising  of a periodic modulation of the crystal, the modulation period  being larger than the lattice
atomic spacing. The new ordering, which is  usually called charge density wave, opens a gap in the electron band
at the initial Fermi energy. In this case the Fermi energy decreases together with a decrease of the total energy of
electrons that stabilizes the charge density wave, compensating the increase of the energy contribution associated with
the source of a mechanism responsible for the periodic modulation, like the coupling of band electrons to phonons or
some other boson field, electron-electron interaction, etc.

In two dimensional (2D) anisotropic  materials the formation of density waves (DWs) is also possible if the highly
anisotropic Fermi surface (FS) has  parts of its contour which can be well enough nested, that is, if one part of it can
be mapped onto another one by a single wave vector\cite{Gruner}. The typical examples of such anisotropic materials are
Bechgaard salts \cite{Pouget} which have open FSs with inflection points on the opposite contours coupled by the DW wave
vector. Density waves of this type have been intensely investigated and widely observed in many other anisotropic materials
as well \cite{Gruner,Pouget}.

Besides the above-mentioned cases, DWs have been also observed in many conductors with closed FSs not satisfying the
nesting condition. Among them are underdoped high-$T_{c}$ cuprates \cite{cuprates} and  graphite intercalates \cite{graphene}.
Their FSs are rather isotropic, thus completely excluding the nesting as a mechanism of the DW stabilization. Although such
materials have been intensely investigated, the origin of these structural instabilities is still unclear.

\begin{figure}
\centerline{\includegraphics[width=8.0cm]{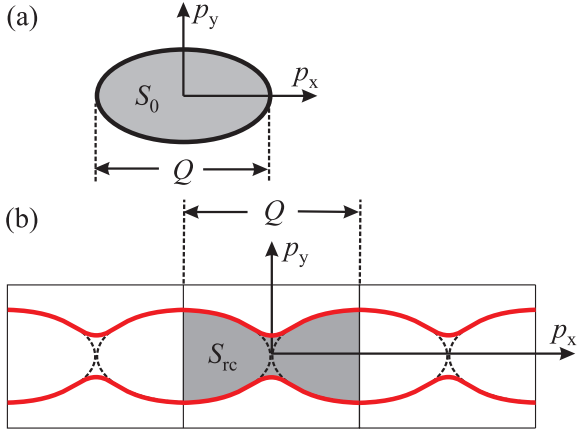}}
\caption{Schematic presentation of the topological transformation of the Fermi surface caused by the charge ordering with
the momentum $Q= 2 p_F$. After lifting the degeneracy at the touching points, initially closed Fermi surfaces in the extended
reciprocal space shown in figure (a), are transformed into an open one with the period $Q$ (red solid lines in figure (b)).
The area of the latter, $S_{rc}$, is larger than the initial Fermi surface $S_0$ at the same energy (see shaded areas). As the
electron number is conserved, the Fermi energy and hence the total band energy decrease, thus stabilizing the charge ordering.}
\label{FSreconstruction}
\end{figure}

In the present paper we suggest a qualitatively new mechanism of the DW ordering, based on a topological reconstruction of
the FS induced by the self-consistently stabilized DW periodic modulation. The initial closed FS is then transformed into an
open one as it is shown in Fig. \ref{FSreconstruction}. We show that this FS reconstruction decreases the electron
band energy, allowing stabilization of the DW by compensating the increase of the energy associated  with the periodic modulation.

Section II contains heuristic considerations leading to the qualitative arguments in favor of the stabilization of the DW ground state
accompanied by the topological reconstruction of the band spectrum. In Sec. III we determine the details of the electron
spectrum in the presence of finite uniaxial modulation $V(x)$. The self-consistent determination of $V(x)$, of the corresponding
total energy of the DW ground state, and of the conditions for its stabilization are discussed in Sec. IV. The concluding
remarks are presented in Sec. V.

\bigskip
\bigskip

\textbf{II. Qualitative considerations}

\bigskip
\bigskip

We consider, as an illustrative example, a 2D conductor which initially has a simple quadratic band dispersion,
$\varepsilon(k_{x}, k_{y}) = (k_{x}^{2} + k_{y}^{2})/2m$. Here $m$ is the electron effective mass. Let us introduce within a
mean-field scheme an uniaxial periodic charge modulation in the $x$-direction which causes a DW potential
$V(x)=\Delta \cos{(Qx/\hbar + \Phi)}$. $Q$ and $\Delta$ are the momentum and the amplitude of the DW order parameter respectively. They will be determined self-consistently by the minimization of the total energy in the Section IV. The phase $\Phi$ of DW potential  will not be important in our considerations since we calculate only the ground state. We start with
the assumption, which will be confirmed by further analysis, that the value of the momentum $Q$ is equal, or close, to the
doubled Fermi momentum $p_{F0}=\sqrt{2m \varepsilon_{F0}}$, where $\varepsilon_{F0}$ is the initial Fermi energy. The
potential $V(x)$ with such modulation combines initially closed FSs in the extended reciprocal space into an
infinite chain of FSs with lifted degeneracy at the touching points, and with a new first Brillouin zone defined, after the
change of coordinates in the reciprocal space
$p_{x} = k_{x} + Q/2, p_{y} = k_{y}$, by $-Q/2 \leq p_{x} \leq Q/2$, as it is shown in Fig. \ref{FSreconstruction}.

To get an initial insight into this band reconstruction, let us provisionally choose $Q = 2 p_{F0}$, although the true
equilibrium value of $Q$ will be modified, as it is shown in Appendix A. As one readily sees in Fig. \ref{FSreconstruction}. a,
the area within the reconstructed Fermi contour inside one cell of  the new reciprocal lattice,  $S_{rc}(\varepsilon_{F0})$,
is larger than the area  $S_{0}(\varepsilon_{F0})$ of the initial closed FS. Therefore, the number of states
$n(\varepsilon_{F0})=S_{rc}(\varepsilon_{F0})/(2\pi \hbar)^2$ at the same energy would be larger than the initial one, and
hence the Fermi energy should decrease to equalize the areas in order to  keep the number of electrons unchanged,  $n_{rc}(\varepsilon_{F})=n_{0}(\varepsilon_{F0})$, i. e.
\begin{eqnarray}
\int_{\varepsilon(p_x,p_y)=\varepsilon_F}dp_x dp_y = \pi p^2_{F0},
\label{elnumbconserve}
\end{eqnarray}
where integration is over the area of the new FS. The decrease of the Fermi energy after such a reconstruction is
accompanied by a decrease of the electron band energy according to the extended theorem of small increments \cite{Landau}.
This decrease can stabilize the DW by compensating the increase of the contribution to the total energy due to the formation
of the periodic potential $V(x)$, as already indicated before.

\begin{figure}
\centerline{\includegraphics[width=8.0cm]{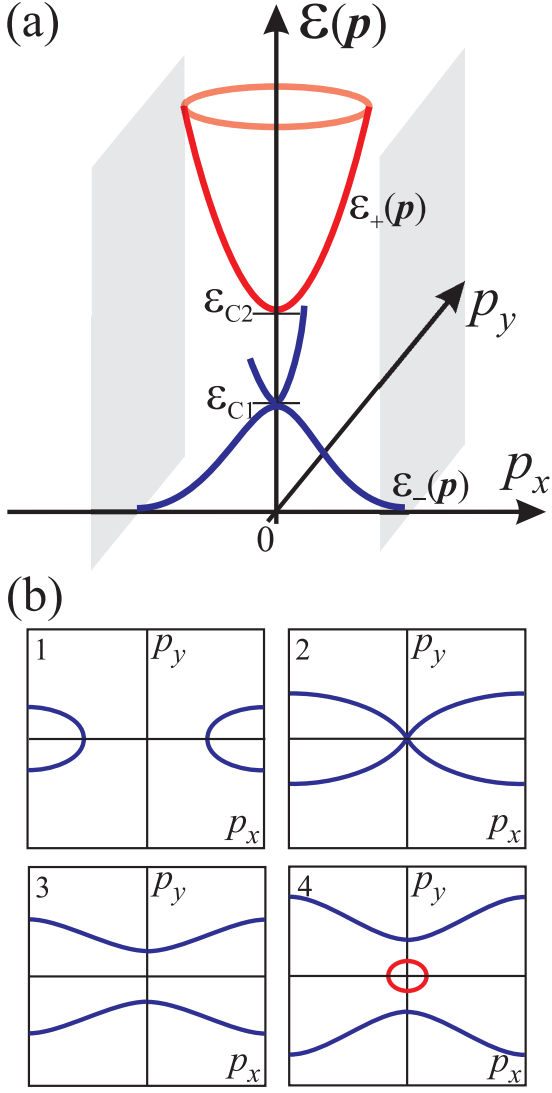}}
\caption {\textbf{(a)} Schematic presentation of topological reconstruction of the spectrum of  a 2D electron gas
(following from Eq.(\ref{NewBadsTransform})) in the vicinity of the saddle point
$\varepsilon=\varepsilon_{C1} =(Q/2)^2/2m - \Delta$ inside one cell of the new Brillouin zone. Energy
$\varepsilon=\varepsilon_{C2}=(Q/2)^2/2m + \Delta $ is the bottom of the  new energy band.
\textbf{(b)} Equienergetic lines in the $(p_x,p_y)$-plane for: energy below (1) and at the saddle point $\varepsilon_{C1}$ (2);
energy between the saddle point and the bottom  of the upper band, $\varepsilon_{C1} < \varepsilon <\varepsilon_{C2}$ (3); and  energy above the bottom of the upper band, $\varepsilon > \varepsilon_{C2}$, when the closed pocket of the upper band appears (4).}
\label{2Ddispersion}
\end{figure}

On the other hand, one sees that at a large enough FS overlapping (caused by a decrease of $Q$) the total area
remains  nearly the same, or even decreases, after lifting the degeneracy at the crossing points. Hence the band
energy is nearly the same as the initial one. From here and the above considerations it follows that the band
energy of the reconstructed system has a minimum in the vicinity of the touching points, i. e. for $Q$ close to
$2 p_{F0}$. Our analytical calculations presented below, as well as the detailed insight into the density of states,
confirm the decrease of the energy caused by this topological reconstruction of the FS. They also show that the
band energy has a minimum when the new Fermi energy is slightly below the upper critical energy $\varepsilon_{C2}$,
at which the new upper band $\varepsilon_{+}(\mathbf{p})$ appears (see Fig. \ref{2Ddispersion}), obeying the condition
$\varepsilon_{F}=\varepsilon_{F0}$. Note that the above-mentioned theorem \cite{Landau} of the small increments
can not be used while considering the minimum of one of the thermodynamic potentials under variations of the parameter.

\bigskip
\bigskip

\textbf{III. Topological reconstruction of the electron band}

\bigskip
\bigskip

Here we consider the DW ground state at the temperature $T=0$ using for the sake of definiteness the
standard Fr\"ohlich electron-phonon Hamiltonian
\begin{eqnarray}
H=\sum_{\mathbf{k}}\varepsilon(\mathbf{k})a^{\dag}_\mathbf{k}a_\mathbf{k} +
\sum_{\mathbf{q}}\hbar \omega(\mathbf{q})b^{\dag}_\mathbf{q}b_\mathbf{q}\nonumber \\ +
\frac{1}{\sqrt{A}}g\sum_{\mathbf{k},\mathbf{q}}
a^{\dag}_{\mathbf{k}+\mathbf{q}}a_\mathbf{k}\left(b^{\dag}_{-\mathbf{q}}+b_{\mathbf{q}}\right),
\label{Froehlich}
\end{eqnarray}
although our reasoning can be extended to physical cases of band electrons coupled to some other boson field,
or through some mutual electron-electron interaction. Here $A$ is the area of the two-dimensional system,
$a^{\dag}_\mathbf{k}$, $a_\mathbf{k}$ and $b^{\dag}_\mathbf{q}$, $b_\mathbf{q}$ are the creation and annihilation
operators for electron states with energy $\varepsilon(\mathbf{k})$ and momentum $\mathbf{k}=(k_x,k_y)$, and phonon
states with energy $\hbar \omega (\mathbf{q})$ and momentum $\mathbf{q}=(q_x,q_y)$, respectively. $g$ is the
electron-phonon coupling constant, for simplicity assumed to be independent of momenta $\mathbf{k}$ and
$\mathbf{q}$.

After assuming the presence of a finite DW modulation, and treating it within the mean-field
approximation, the above Hamiltonian reduces to its mean-field form,
\begin{eqnarray}
H_{MF}=\sum_\mathbf{k}\left[ \varepsilon(\mathbf{k})a^{\dag}_\mathbf{k}a_\mathbf{k} +
\Delta e^{i \Phi} a^{\dag}_{\mathbf{k+Q}}a_\mathbf{k}+ \Delta e^{-i \Phi}
a^{\dag}_{\mathbf{k-Q}}a_\mathbf{k}\right] \nonumber \\ + \frac{A \hbar \omega_{Q}}{2 g^{2}} \Delta^{2},
\hspace{0.5cm}
 \label{meanfieldH}
\end{eqnarray}
 where
\begin{eqnarray}
\sqrt{A} \Delta e^{i \Phi} =g \left(\langle b_\mathbf{Q} \rangle + \langle b_\mathbf{-Q}^\dag
\rangle\right)
 \label{Delta}
\end{eqnarray}
is the order parameter, $\langle b_\mathbf{Q} \rangle=\langle b_\mathbf{-Q}^\dag \rangle$
is the non-vanishing expectation value of macroscopically occupied DW phonon mode.
The values of the order parameter $\Delta$ and the DW momentum $Q$ will be determined later
by the minimization of the total energy of the system.

After the diagonalization of the electronic part of Hamiltonian (\ref{meanfieldH}) one finds the
electron spectrum of the perturbed system as follows:
\begin{eqnarray}
\varepsilon_{\pm}(\mathbf{p})= \frac{\varepsilon_{1}(\mathbf{p})+\varepsilon_{2}(\mathbf{p})}{2}\pm
 \sqrt{\Big(\frac{\varepsilon_{1}(\mathbf{p})-\varepsilon_{2}(\mathbf{p})}{2} \Big)^2+\Delta^2},
\label{NewBands}
\end{eqnarray}
where
\begin{eqnarray}
 \varepsilon_{1,2}(\mathbf{p})\equiv \varepsilon(p_x \pm Q/2, p_y)=\frac{\left(p_x\pm Q/2\right)^2}{2m}+\frac{p^2_y}{2m},
\label{ShifDispersion}
\end{eqnarray}
is presented in terms of the momenta of new Brillouin zone introduced in Sec. II. The gap in the electron spectrum
$\Delta$ in Eq.(\ref{NewBands}) is defined in Eq.(\ref{Delta}). With the expression (\ref{ShifDispersion}) taken into account,
the spectrum (\ref{NewBands}) reads
\begin{eqnarray}
\varepsilon_{\pm}(\mathbf{p})=\frac{(Q/2)^2 +p_y^2 +p_x^2 \pm \sqrt{(Qp_x)^2+(2m \Delta)^2}}{2m}.
\label{NewBadsTransform}
\end{eqnarray}

The dispersions and constant energy surfaces (CESs) of the new electron bands $\varepsilon_{\pm}(\mathbf{p})=\varepsilon$
are shown in Fig. \ref{2Ddispersion}. One sees that CESs  of the lower band $\varepsilon_{-}(\mathbf{p})= \varepsilon$
are present at all energies above the bottom of the original band (slightly lowered due to the contribution of the order of $\Delta^{2}/2\varepsilon_{F0}$). This band has a saddle point at the wave vector $p_{x} = p_{y} = 0$ and the energy
$\varepsilon_{C1}=(Q/2)^2/2m -\Delta $. The upper band $\varepsilon_{+}(\mathbf{p})= \varepsilon$ is bounded from below, with
the bottom at $p_{x} = p_{y} = 0$ and the energy $\varepsilon_{C2}=(Q/2)^2/2m +\Delta $. In the next Section we show that
these peculiar topological properties of the reconstructed electron band structure result in a decrease of the
total band energy and a possible occurrence of the DW.

\bigskip
\bigskip

\textbf{IV. Band energy and the stabilization of the DW}

\bigskip
\bigskip

Since the detailed analysis confirms the qualitative arguments from Sec. II about the regime in which the DW stabilization
could take place, we limit further considerations to the range of values of the momentum Q for which the the initial
Fermi energy  $\varepsilon_{F0}$ is between the saddle point of the lower band, $\varepsilon_{C1}$,  and the minimum of the
upper band, $\varepsilon_{C2}$. In this range the total electron band energy per unit area for the reconstructed system is

\begin{eqnarray}
 E_B(\varepsilon_{F})= 2 \int_{\varepsilon_{-} (\mathbf{p}) =
 \varepsilon_F}\varepsilon_{-} (\mathbf{p}) \frac{d^{2} \mathbf{p}}{(2 \pi \hbar)^2} =\nonumber \\
  \frac{4}{(2\pi \hbar)^2 m}        \int_{0}^{Q/2}d p_x
\int_{0}^{p_y^{(F-)}}d p_y    \hspace{1.5 cm}    \nonumber \\
\times \left[\Big( \frac{Q}{2}\Big)^2+p_y^2+p_x^2-\sqrt{(Q p_x)^2+ (2m \Delta)^2}\right],\;
 \label{BandEnergy}
\end{eqnarray}

where the factor 2 comes from the spin degeneracy, and
\begin{eqnarray}
p_y^{(F-)}(p_x)= \nonumber \\
\Big\{2m\varepsilon_{F}-\left(\frac{Q}{2}\right)^2-p_x^2+\sqrt{(Q p_x)^2+(2m \Delta)^2}\Big\}^{1/2}.
\label{py}
\end{eqnarray}

Using Eq.(\ref{py}), and subtracting the initial band energy
\begin{eqnarray}
E_0=  2 \pi m \varepsilon_{F0}^2/ (2\pi \hbar)^2
\label{E0}
\end{eqnarray}
from Eq.(\ref{BandEnergy}), one finds the decrease of the total band energy per unit area as follows:
\begin{eqnarray}
\Delta E_{B} \equiv E_{B}-E_0 = \frac {4 \pi m}{(2 \pi \hbar)^2}\Big\{\varepsilon_{F}\varepsilon_{F0}-\frac{\varepsilon_{F0}^2}{2} \nonumber \\
-\frac{8}{3 \pi} \frac{1}{(2 m)^2} \int_0^{Q/2} \Big[p_y^{(F-)}(p_x)\Big]^3  d p_x\Big\}.
\label{energydecrease}
\end{eqnarray}

The Fermi energy $\varepsilon_F$ of the reconstructed system is determined from the condition (\ref{elnumbconserve}) by which
the band reconstruction does not change the number of electrons. It can be rewritten as
\begin{eqnarray}
4 \int_{0}^{Q/2} p_y^{(F-)}(p_x; \varepsilon_F, Q) d p_x =2\pi m \varepsilon_{F0}.
\label{conservenumber2}
\end{eqnarray}

Equations (\ref{py}), (\ref{energydecrease}) and (\ref{conservenumber2}) determine the dependence of the decrease of the electron
band energy with the initial Fermi energy $\varepsilon_{F0}$ on the momentum $Q$ and the amplitude $\Delta$ of the DW. The optimal
values od $Q$ and $\Delta$ follow from the minimization of this energy decrease, with the condition (\ref{conservenumber2})
taken into account. Also, the usually weak $Q$-dependence of the factor in front of $\Delta^2$ within the last term in $H_{MF}$, Eq.(\ref{meanfieldH}), is neglected as
nonessential for the key qualitative conclusions to be drawn here.

The first question to be answered is: given the value of the DW order parameter $\Delta$, what is the value of the DW wave momentum,
$Q_{m}$, which minimizes the total band energy (\ref{energydecrease})? The minimization performed in the Appendix A. It leads to the
conclusion that this momentum is determined by the condition
\begin{eqnarray}
\varepsilon_{F}=\varepsilon_{F0}.
\label{optimalepsilon}
\end{eqnarray}
In other words, the optimal DW order takes place when the Fermi energies of the reconstructed and initial bands coincide.
The corresponding value of the momentum $Q_{m}$ as a function of the DW amplitude $\Delta$ follows from the relation
(\ref{conservenumber2}), with the condition (\ref{optimalepsilon}) inserted.

Furthermore, inserting this condition also into the expressions (\ref{energydecrease}), one can perform the minimization of
the total energy of the DW with the conservation of electrons taken into account, and with the lattice part $E_{\emph{latt}}$ originating
from the last term in the Hamiltonian $H_{MF}$ (Eq. \ref{meanfieldH}) included,
\begin{eqnarray}
\frac{E_{DW}}{E_0} = \frac{\Delta E_{B}+ E_{\emph{latt}}}{E_0} =  \nonumber \\
1 - \frac{16}{3\pi p_{F0}^{4}} \int_0^{Q/2} \Big[p_y^{(F-)}(p_x)\Big]^3  d p_x
+ \frac{1}{\lambda}\frac{\Delta^{2}}{\varepsilon_{F0}^{2}}.
\label{totalinitial}
\end{eqnarray}
Here
\begin{eqnarray}
\lambda \equiv \frac{m}{\pi\hbar^{2}}\frac{g^{2}}{2\hbar\omega_{Q}}
\label{lambdadef}
\end{eqnarray}
 is the usual definition of the dimensionless
electron-phonon coupling constant. Note that $m/(\pi\hbar^{2})$ is the density of states of the initial two-dimensional electron band.
\begin{figure}
\centerline{\includegraphics[width=8.0cm]{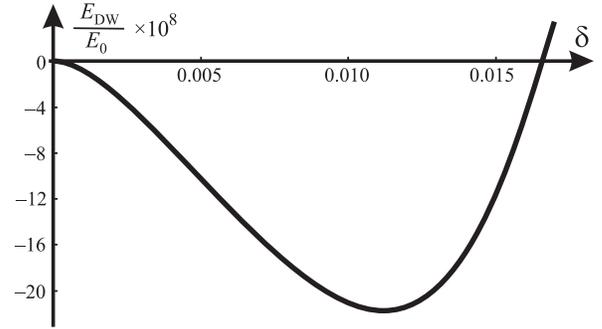}}
\caption{Dependence of the normalized total energy change
$\frac{E_{DW}}{E_0}$ due to the DW order on the value of normalized order parameter $\delta$ for the value of dimensionless
coupling constant  $\lambda=0.613$. }
\label{totalenergy}
\end{figure}

The numerical minimization of eqs.(\ref{conservenumber2}, \ref{totalinitial}) leads to the dependence of the total energy
$E_{DW}/{E_0}$ on the normalized order parameter $\delta \equiv \Delta/\varepsilon_{F0}$ shown in Fig. \ref{totalenergy},
indicating the stabilization of the ordered state with a finite value of the DW amplitude. The more direct analytical insight into
the characteristics of this ordering follow from the expansions of the expressions (\ref{conservenumber2}, \ref{totalinitial})
which reproduces to a high level of accuracy the numerical results.

The first step in the analytical analysis is the determination of the momentum $Q_{m}$ of the stabilized DW order as a
function of the order parameter $\delta$ given by the Eq. (\ref{conservenumber2}) with $\varepsilon_{F}=\varepsilon_{F0}$. The
expansion performed in the Appendix A leads to the result
\begin{eqnarray}
q_{m} \equiv \frac{Q_m }{2 p_{F0}} \approx 1- \frac{\delta}{2}  + \frac{1}{\pi} \Big(\frac{\delta}{2}\Big)^{3/2} + \textsl{O}(\delta^{2}).
\label{optimalQ}
\end{eqnarray}

With the known dependence of $Q_{m}$ on $\delta$, one can perform the second step, elaborated in the Appendix B, the expansion
of the total energy (\ref{totalinitial}) in terms of the order parameter $\delta$. The leading terms are given by
\begin{eqnarray}
\frac{E_{DW}}{E_0} = \left(\frac{1}{\lambda} -
\frac{1}{\lambda_c}\right)\delta^{2} + \frac{1}{\pi}\delta^3 + \textsl{O}(\delta^{7/2}),
\label{Expandtotal}
\end{eqnarray}
where the last term indicates the lowest possible order of power not covered by the expansion procedure in the Appendix B.
The critical value of the coupling constant $\lambda$ is given by
\begin{eqnarray}
\lambda_c= \frac{1}{1+\frac{2}{\pi}}.
\label{lambdac}
\end{eqnarray}

\begin{figure}
\centerline{\includegraphics[width=8.0cm]{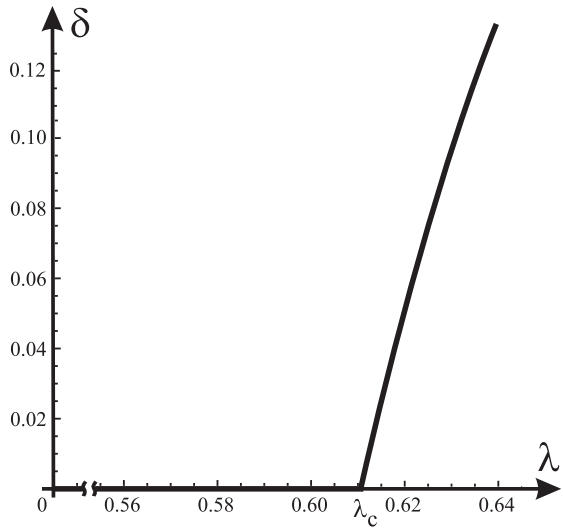}}
\caption{ Dependence of the normalized order parameter
$\delta$ on the dimensionless
coupling constant $\lambda$. Here, $\lambda_c=(1+2/\pi)^{-1}$  is the critical point of the quantum phase
transition at which the homogeneous state $\delta =0$ looses its stability and the stable DW ground state
with $\delta \neq 0$ is formed.}
\label{lambda}
\end{figure}

Minimization of the total energy (\ref{Expandtotal}) leads to the equation for $\delta_{m}$, the equilibrium
value of the normalized order parameter,
\begin{eqnarray}
2\delta \Big(\frac{1}{\lambda} - \frac{1}{\lambda_c} +  \frac{3}{2\pi} \delta \Big) = 0.
\label{gap}
\end{eqnarray}
The nontrivial and stable solution is given by
\begin{eqnarray}
\delta_{m}=\frac{2\pi}{3} \Big(\frac{1}{\lambda_{c}} -\frac{1}{\lambda}\Big).
\label{OrderParamenter}
\end{eqnarray}
It appears in the range of values of the  coupling constant  $\lambda \geq \lambda_{c}$,
while below this critical value  the stable  solution is that for the non-ordered state, $\delta = 0$. The
latter solution looses its stability at $\lambda \geq \lambda_c $, as shown in Fig.\ref{lambda}. Hence $\lambda_{c}$
defines the critical point of the quantum phase transition.

We conclude this analysis of the energy of the DW ordering with four notes.

First, the expansion of the band energy change due to its topological reconstruction, $\Delta E_{B} \equiv E_{B}-E_0$ given by
Eq. (\ref{energydecrease}), shows that it has a minimum as a function of the amplitude of the order parameter $\delta$, and
changes sign as $\delta$ further increases, in contrast to the case of conventional DW orderings in which it remains negative for
all values of $\delta$. Furthermore the expansion of the total energy change (\ref{Expandtotal}) differs from that of the
standard continuous (second order) phase transition, since the stabilizing contribution is cubic, and not quartic, in the
order parameter amplitude $\delta$.

Second, since the decrease of the band energy is quadratic in $\delta$, in contrast to usual nesting instabilities in which
it shows divergent tendency as $\delta \rightarrow 0$, the DW instability is possible only provided the coupling constant
$\lambda$ is large enough, as the result (\ref{Expandtotal}, \ref{lambdac}) shows.

Third, the momentum of the DW modulation varies with the order parameter $\delta$, again in contrast to the conventional
low dimensional DW cases in which it is fixed by the geometric nesting constraint. This is one of the reasons for our
inclination to term the present mechanism the \emph{touching}, and not the \emph{nesting}, one, in accordance with the geometry from
Fig. \ref{FSreconstruction}.

Fourth, as it is seen from Fig. \ref{lambda}, the order parameter $\delta$ steeply increases with $\lambda - \lambda_{c}$,
which limits the validity of our mean-field approach, and the ensuing expansions for the DW momentum and energy, to a rather narrow
range close to the quantum critical point. With  $\delta$ approaching the value of unity (i. e. with the gap $\Delta$ in the
reconstructed band dispersion from Fig. \ref{2Ddispersion} becoming of the order of the Fermi energy) the whole present
approach has to be replaced by a more rigorous treatment.

\bigskip

\textbf{V. Conclusions}

\bigskip

We have shown that an isotropic 2D system with closed Fermi surface (i. e. with entirely
excluded nesting conditions) may be topologically reconstructed due to the stabilization of uniaxially
modulated DW. By this reconstruction the initially closed Fermi surface is transformed into an open one
in the extended reciprocal space as it is presented in Fig.\ref{FSreconstruction}. Such a topological transformation
of the Fermi surface  decreases the electron band energy, enabling the stabilization of the DW.
More precisely, we have found that the DW is stable if the coupling constant is larger than the critical one,
$\lambda \geq \lambda_c= (1+2/\pi)^{-1}$. With this condition fulfilled, the system  undergoes the quantum phase transition
under a change of the parameter $\lambda$ as it is shown in Fig. \ref{lambda}. The obtained values of order parameter $\Delta$
can exceed $10^{-2 } \varepsilon_{F}$, that is the critical temperature of the phase transition can be $T_c \sim 10^{2}$K.

The above qualitative and quantitative proposals indicate that the concept of the topologically reconstructed FSs
invoked in the present work, may be the source of  the  density waves frequently observed in 2D conductors such as
high-$T_c$ cuprates and graphite intercalates. However, for the more detailed quantitative explanations of the
phase diagram for these materials, it is necessary to take into account specific geometries and dispersions in their band
structures.

The further question deserving future analysis is the behavior of the reconstructed spectrum from
Fig. \ref{FSreconstruction} under strong magnetic fields. Having the coexistence of open and closed orbits and the barriers
between them, one meets the possibility of an additional gain in the band energy due to the effect of magnetic
breakdown \cite{KaganovSlutskin}, already encountered in such \cite {PhysB}, or similar \cite{PRL,EPJB}, band spectra.
Preliminary analyzes indeed confirm that such energy gains take place, as will be elaborated in our forthcoming paper.

\textbf{\emph{Acknowledgement}}. This work was supported by the Croatian Science Foundation, project IP-2016-06-2289,
and by the QuantiXLie Centre of Excellence, a project cofinanced by the Croatian Government and European Union
through the European Regional Development Fund - the Competitiveness and Cohesion Operational Programme
(Grant KK.01.1.1.01.0004).

\bigskip

\textbf{Appendix A: The optimal value of the DW momentum}

\bigskip

In this Appendix we determine the value of the momentum $Q = Q_{m}$ which minimizes the total band energy
(\ref{energydecrease}) for a given, presumably finite, value of the order parameter $\Delta$. Among a few possible ways
leading to the result, that performed below appears to be the simplest.

Let us assume that the momentum $Q$ is fixed, and look for the band filling, i. e. for the value of the Fermi
energy $\varepsilon_{F0}$, for which the total energy of the reconstructed band $\Delta E_{B}$ is minimal, i. e. for which
the condition
\begin{eqnarray}
\frac{d\Delta E_{B}}{d\varepsilon_{F0}} = \frac{\partial\Delta E_{B}}{\partial\varepsilon_{F0}} + \frac{\partial\Delta E_{B}}{\partial\varepsilon_{F}}\frac{\partial\varepsilon_{F}}{\partial\varepsilon_{F0}} = 0
\label{epsilon0minim}
\end{eqnarray}
is satisfied. Here we take into account that $\varepsilon_{F}$ appearing in Eq.(\ref{energydecrease}) depends on
$\varepsilon_{F0}$ through the expression (\ref{conservenumber2}). After performing the derivatives the condition
(\ref{epsilon0minim}) reduces to
\begin{eqnarray}
\frac{d\Delta E_{B}}{d\varepsilon_{F0}} =  \frac{4 \pi m}{(2 \pi \hbar)^2} (\varepsilon_{F} - \varepsilon_{F0}) = 0,
\label{epsiloncondition}
\end{eqnarray}
i. e. the band energy gain (\ref{energydecrease}) is maximal when the Fermi energies of the reconstructed band and the
initial band coincide. The condition (\ref{epsiloncondition}) is the local minimum  of $\Delta E_{B}$ if
$d^{2}\Delta E_{B}/d\varepsilon_{F}^{2} > 0$, i. e. for
\begin{eqnarray}
\Big[\frac{\partial \varepsilon_{F}}{\partial \varepsilon_{F0}}\Big]_{\varepsilon_{F}=\varepsilon_{F0}} > 1.
\label{mimimum}
\end{eqnarray}

$Q_{m}$, the momentum which minimizes $\Delta E_{B}$,  as a function of $\varepsilon_{F0}$ now follows from the equality
(\ref{conservenumber2}) with the condition $\varepsilon_{F} = \varepsilon_{F0}$ inserted into its left-hand side. Before
deriving the approximative solution, it is useful to introduce dimensionless variables, with scales absorbing the momentum
$Q_{m}/2$,
\begin{eqnarray}
\frac{2m\varepsilon_{F0}}{(Q_{m}/2)^{2}} = \frac{1}{q_{m}^{2}}\equiv \widetilde\varepsilon_{F0}, \,\,\,\,\,
\nonumber \\  \frac{2m \Delta}{(Q_{m}/2)^{2}} = \frac{\delta}{q_{m}^{2}} \equiv \widetilde{\delta}, \,\,\,\,\,
\frac{p_{x}}{Q_{m}/2}\equiv x.
\label{transformation}
\end{eqnarray}
Eq. (\ref{conservenumber2}) then reads
\begin{eqnarray}
\int_{0}^{1} dx \sqrt{\widetilde\varepsilon_{F0} - 1 - x^{2} + \sqrt{4x^{2} + \widetilde{\delta}^{2}}} = \frac{\pi}{4} \widetilde\varepsilon_{F0}.
\label{dimensionless}
\end{eqnarray}
The numerical insight into this equation indicates that in the physical range of values of order parameter,
$\widetilde{\delta} \ll 1$, the solution for $\varepsilon_{F0}$ is slightly below the critical value $\varepsilon_{C2}$
(i. e. $\widetilde\varepsilon_{F0}$ is slightly below $1 + \widetilde{\delta}$ in terms of dimensionless variables (\ref{dimensionless})).
We therefore write
\begin{eqnarray}
\widetilde\varepsilon_{F0} - (1 + \widetilde{\delta}) \equiv  \emph{f}(\widetilde{\delta}),
\label{epsilonexp}
\end{eqnarray}
and expand the left-hand side of the equality (\ref{dimensionless}) in terms of presumably small difference $\emph{f}(\widetilde{\delta})$.
The straightforward calculation leads to the result for the leading term
\begin{eqnarray}
\emph{f}(\widetilde{\delta}) \simeq - \frac{1}{\sqrt{2}\pi} \widetilde{\delta}^{3/2},
\label{f}
\end{eqnarray}
which is consistent with the above initial assumption. Inserting original physical variables into expressions
(\ref{epsilonexp}, \ref{f}), we finally get the expansion for the momentum $q_{m}$ given by Eq.(\ref{optimalQ}).

\begin{figure}
\centerline{\includegraphics[width=8.0cm]{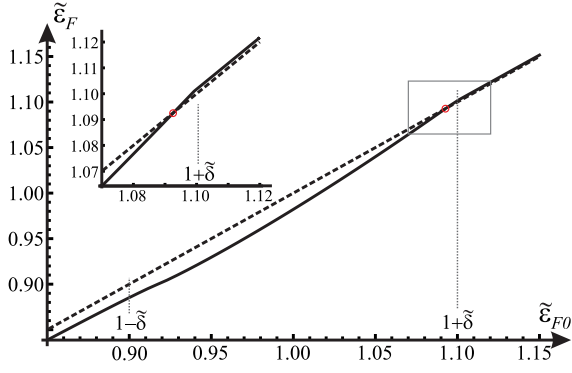}}
\caption{The Fermi energy of the reconstructed band $\widetilde\varepsilon_{F}$ as the function of the original Fermi energy $\widetilde\varepsilon_{F0}$ for
$\delta = 0.1$. The red point marks the equilibrium condition $\widetilde\varepsilon_{F} = \widetilde\varepsilon_{F0}$, Eq. (\ref{optimalepsilon}).}
\label{eF}
\end{figure}

The obtained analytical results complement and quantitatively confirm the heuristic considerations from Sec. II. They are also in
full agreement with numerical calculations shown in Fig. \ref{eF}.  The crossing of the lines
$\widetilde\varepsilon_{F}(\widetilde\varepsilon_{F0})$ and $\widetilde\varepsilon_{F}= \widetilde\varepsilon_{F0}$ is
indeed realized slightly below the bottom of the upper
sub-band from Fig.(\ref{2Ddispersion}) positioned at the energy $  \widetilde\varepsilon\equiv 2m\varepsilon / (Q_{m}/2)^{2} = 1 + \widetilde{\delta}$.
Furthermore, $\widetilde\varepsilon_{F}(\widetilde\varepsilon_{F0})$ crosses the line $\widetilde\varepsilon_{F}= \widetilde\varepsilon_{F0}$
from below by approaching it from the
left, which guaranties the fulfilment of the condition (\ref{mimimum}). Note also that the minimum of the band energy
$\Delta E_{B}$ is realized in the range of reconstructed band filling in which the Fermi level is within the lower sub-band,
although very close to the bottom of the upper sub-band. Since the latter does not contribute to the band energy, we did not
have to include it into the present analysis.

\bigskip

\textbf{Appendix B: The expansion of the total energy}

\bigskip

In this Appendix we minimize the total energy (\ref{totalinitial}), and expand it in terms of the order parameter $\delta$. To this end,
it appears convenient to use again the reduced quantities from the previous Appendix. Introducing them into its band part,
the total energy is given by
\begin{eqnarray}
\frac{E_{DW}}{E_{0}} =
1 - \frac{16}{3 \pi } q_{m}^{4} I + \frac{1}{\lambda}\delta^{2},
\label{red-energy}
\end{eqnarray}
with
\begin{eqnarray}
I \equiv \int_0^{1} \Big[q_{m}^{-2} - 1 -  x^{2} + \sqrt{4 x^{2} + \widetilde{\delta}^{2}}\Big]^{3/2}  d x.
\label{I}
\end{eqnarray}
The conservation of electrons (\ref{conservenumber2}) now reads
\begin{equation}
q_{m}^{2} \int_0^{1} \Big[q_{m}^{-2} - 1- x^{2} + \sqrt{4 x^{2} + \widetilde{\delta}^{2}}\Big]^{1/2} d x = \pi/4.
\label{cons}
\end{equation}

Let us in the first step consider the derivative $d I/d \widetilde{\delta}$. It is given by
\begin{equation}
\frac{d I}{d \widetilde{\delta}} = \frac{\partial I}{\partial q_{m}^{-2}} \frac{d q_{m}^{-2}}{d \widetilde{\delta}} +
\frac{\partial I}{\partial \widetilde{\delta}}.
\label{dI}
\end{equation}
Since, from eqs. (\ref{I}) and (\ref{cons}) one gets
\begin{equation}
\frac{d I}{d q_{m}^{-2}} = \frac{3\pi}{8 q_{m}^{2}},
\label{dIq}
\end{equation}
Eq. (\ref{dI}) reads
\begin{eqnarray}
\frac{dI}{d\widetilde{\delta}} = \frac{3\pi}{16} \frac{d (q_{m}^{-4})}{d \widetilde{\delta}} \nonumber \\
+ \frac{3}{2} \widetilde{\delta} \int_0^{1} \Big[q_{m}^{-2} - 1- x^{2} + \sqrt{4 x^{2} + \widetilde{\delta}^{2}}\Big]^{1/2} \frac{d x}{\sqrt{4 x^{2} + \widetilde{\delta}^{2}}}.
\label{dIbis}
\end{eqnarray}
Integrating this expression with respect to $\widetilde{\delta}$, and taking into account that for $\widetilde{\delta} = 0$
(and $q_{m}^{-2}=1$, as it follows from eqs. (\ref{transformation}) and (\ref{epsilonexp})) the expression (\ref{I}) reduces to
\begin{equation}
I(\widetilde{\delta}=0) = \frac{3\pi}{16},
\label{I0}
\end{equation}
one gets for the total energy (\ref{red-energy})
\begin{eqnarray}
\frac{E_{DW}}{E_{0}} = - \frac{8 q_{m}^{4}}{\pi} \int_0^{\widetilde{\delta}}  \widetilde{\delta}'  J[\epsilon(\widetilde{\delta}'), \widetilde{\delta}'] d \widetilde{\delta}' + \frac{1}{\lambda}\delta^{2}.
\label{red-energy1}
\end{eqnarray}
where
\begin{eqnarray}
J[\epsilon(\widetilde{\delta}), \widetilde{\delta}] \equiv \int_0^{1} \Big[ \epsilon - x^{2} + \sqrt{4 x^{2} + \widetilde{\delta}^{2}}\Big]^{1/2} \frac{d x}{\sqrt{4 x^{2}   + \widetilde{\delta}^{2}}},
\label{int2}
\end{eqnarray}
with the short-hand notation $\epsilon \equiv q_{m}^{-2} - 1$.

In the next step we expand the integral $J[\epsilon(\widetilde{\delta}), \widetilde{\delta}]$
in terms of $\widetilde{\delta}$. Note that, although both quantities, $\widetilde{\delta}$ and $\epsilon$, are much smaller than unity, the direct expansion in their powers cannot be controlled due to the diverging nature of integrals appearing in the coefficients at the lower integration boundary $x = 0$. Instead, we divide the integration range $0<x<1$ into two subranges, $0<x<r$ and $r<x<1$, with $0 \ll \widetilde{\delta}, \epsilon \ll r \ll \sqrt{\widetilde{\delta}}, \sqrt{\epsilon} \ll 1$, and make adequate approximations in each subrange. More precisely, in the former subrange the term $x^{2}$ is negligible with respect to the square root
$\sqrt{4 x^{2} + \widetilde{\delta}^{2}}$, while in the latter subrange one can neglect in this square root the term $\widetilde{\delta}^{2}$ with respect to $4 x^{2}$. After these approximations the respective expansions in terms of $\epsilon - \widetilde{\delta}$ and $\widetilde{\delta}$ lead to the result
\begin{equation}
\textsl{J}[\epsilon(\widetilde{\delta}), \widetilde{\delta}] \simeq \frac{2 + \pi}{4} - \frac{\epsilon}{4} + \frac{\pi(\epsilon - \widetilde{\delta})}{4\sqrt{2\widetilde{\delta}}},
\label{J}
\end{equation}
with the cancelation of the $r$-dependent contributions. Since the leading term in the difference $\epsilon - \widetilde{\delta}$ is given by the function $\emph{f}(\widetilde{\delta})$ (Eq. \ref{f}), one gets
\begin{equation}
\textsl{J}(\widetilde{\delta}) \simeq \frac{2 + \pi}{4} - \frac{3}{8} \widetilde{\delta}.
\label{Jfinal}
\end{equation}
Inserting this expansion into the expression (\ref{red-energy1}) one gets the final result (\ref{Expandtotal}) for the leading terms in the expansion of the total DW energy.

\end {document}